# Scaling Behavior of Portevin-Le Chatelier Effect

**A. Sarkar***
*Variable Energy Cyclotron Centre*
*1/AF Bidhan Nagar, Kolkata 700064, India*
*E-mail:* apu@veccal.ernet.in

**P. Barat, P. Mukherjee, S. K. Bandyopadhyay**
*Variable Energy Cyclotron Centre*
*1/AF Bidhan Nagar, Kolkata 700064, India*
*E-mail:* pbarat@veccal.ernet.in, paramita@veccal.ernet.in, skband@veccal.ernet.in

The scaling behavior of the Portevin-Le Chatelier (PLC) effect is studied by deforming a substitutional alloy, Al-2.5%Mg and an interstitial alloy, low carbon steel (0.15%C, 0.33%Mn, 0.04%P, 0.05%S, 0.15%Si and rest Iron) at room temperature for a wide range of strain rates. To reveal the exact scaling nature, the time series data of true stress vs. time, obtained during the tensile deformation (corrected for drift due to strain hardening by polynomial fitting method), are analyzed by two complementary methods: the standard deviation analysis and the diffusion entropy analysis. From these analyses we could establish that in the entire span of strain rates, PLC effect showed Levy walk type of scaling property.

* Speaker

## 1. Introduction

The Portevin-Le Chatelier (PLC) effect, also known as jerky flow, denotes a plastic instability, which is related to the discontinuous plastic flow and plastic strain inhomogeneities. It has been observed in many dilute metallic solid solutions including both interstitial [1] and substitutional [2]. The PLC effect is usually undesirable due to its detrimental influences like the loss of ductility and the appearance of surface markings on the specimen.

The microscopic origin of the PLC effect is still a matter of debate. The general consensus explains the origin of the PLC effect as the dynamic interaction between the moving dislocation and the diffusing solute atoms. The mobile dislocations which are carrier of the plastic strain move jerkily between the obstacles provided by the other defects. This microstructural process is denoted as the dynamic strain ageing (DSA) [3-5]. The DSA leads to a negative strain rate sensitivity (SRS) of the flow stress within a certain range of the applied strain rates and temperatures when the mobile dislocations and the solute atoms have comparable mobility. Bands of localized deformation are then formed, in association with stress serrations.

The PLC effect has been extensively studied over the last several decades [3-7] with the goal being to achieve a better understanding of the small-scale processes and of the multiscale mechanisms that link the mesoscale DSA to the macroscale PLC effect. The technological goal is to increase the SRS to positive values in the range of temperatures and strain rates relevant for industrial processes. This would ensure material stability during processing and would eliminate the PLC.

The search for correlation in the space-time distribution of dislocation activity during plastic deformation is a major goal in the dislocation dynamics research. Many papers purport to characterize the space-time organization of dislocation activity in different strain rate and temperature regimes. The flurry of interest from physicists comes from their fascination with the self-similar properties of stress drops together with the development of novel concepts and techniques that may provide proper insights. Investigating statistical properties of the stress-time series and the behavior of the stress fluctuations proved to be a useful approach to study the underlying dynamics of the PLC effect.

For tensile tests at constant applied strain rate different type of PLC instabilities are distinguished according to the spatio-temporal organization of the deformation bands. In polycrystals three generic types of bands: type C, B and A can be distingusihed. At low strain rates type C bands appear at random in the space. At medium strain rates type B bands exhibit an oscillatory or intermittent propagation along the tensile axis. At high strain rates type A bands propagate continuously and smoothly as solitary plastic waves [8-10].

Scaling as a manifestation of underlying dynamics is familiar throughout physics. It has been instrumental in helping scientists gain deeper insights into problems ranging across the entire spectrum of science and technology. Scaling laws typically reflect underlying generic features and physical principles that are independent of detailed dynamics or characteristics of particular models. Scale invariance seems to be widespread in natural systems [11]. Numerous examples of scale invariance properties can be found in the literature like earthquakes, clouds, networks etc. [12-15]. In this paper we study the scaling behavior of the PLC effect.

Since the time of the discovery of the PLC effect, the research on the types of serrations has been restricted almost completely to f.c.c substitutional alloy. Serrations in b.c.c alloy systems, which are caused by interstitially dissolved elements seem mostly devoid of any regular features.

## 2. Experimental

In this work we report the study of detailed scaling behavior of the PLC effect in alloys (substitutional and interstitial) with different crystal structure and microstructure. We have carried out tensile testing on polycrystalline flat Al-2.5%Mg alloy and cylindrical low carbon steel samples. For Al-2.5%Mg alloy samples the gauge length, width and the thickness were 25, 5 and 2.3 mm respectively. For the low carbon steel samples the gauge length and diameter were 25 and 5 mm respectively. Samples were tested in a servo controlled INSTRON (model 4482) machine. All the tests were carried out at room temperature (300 K) and consequently there were only one control parameter: the applied strain rate. To monitor closely its influence on the dynamics of jerky flow, strain rate was varied from $6.03\times10^{-5}$ $sec^{-1}$ to $2.21\times10^{-3}$ $sec^{-1}$. The PLC effect was observed throughout the range. The stress-time response was recorded electronically at periodic time intervals (20 Hz). The stress data taken for analysis were corrected for the strain hardening drift by the method of polynomial fitting.

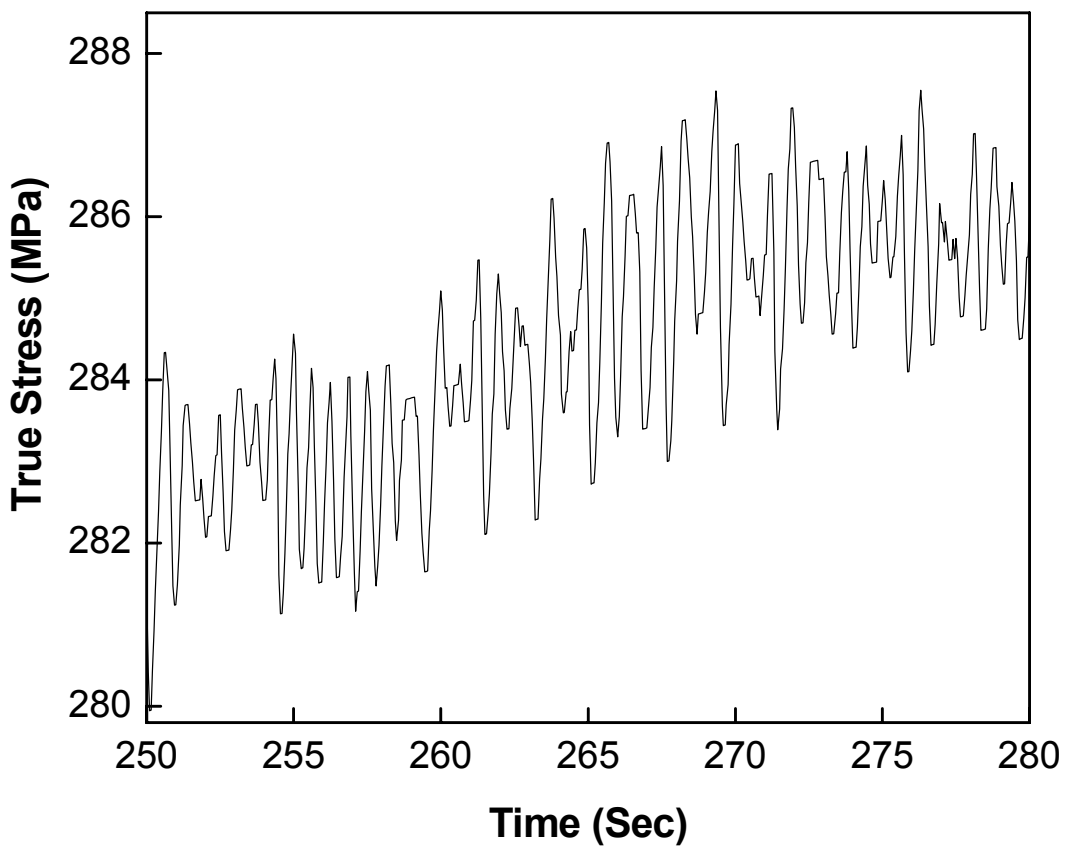

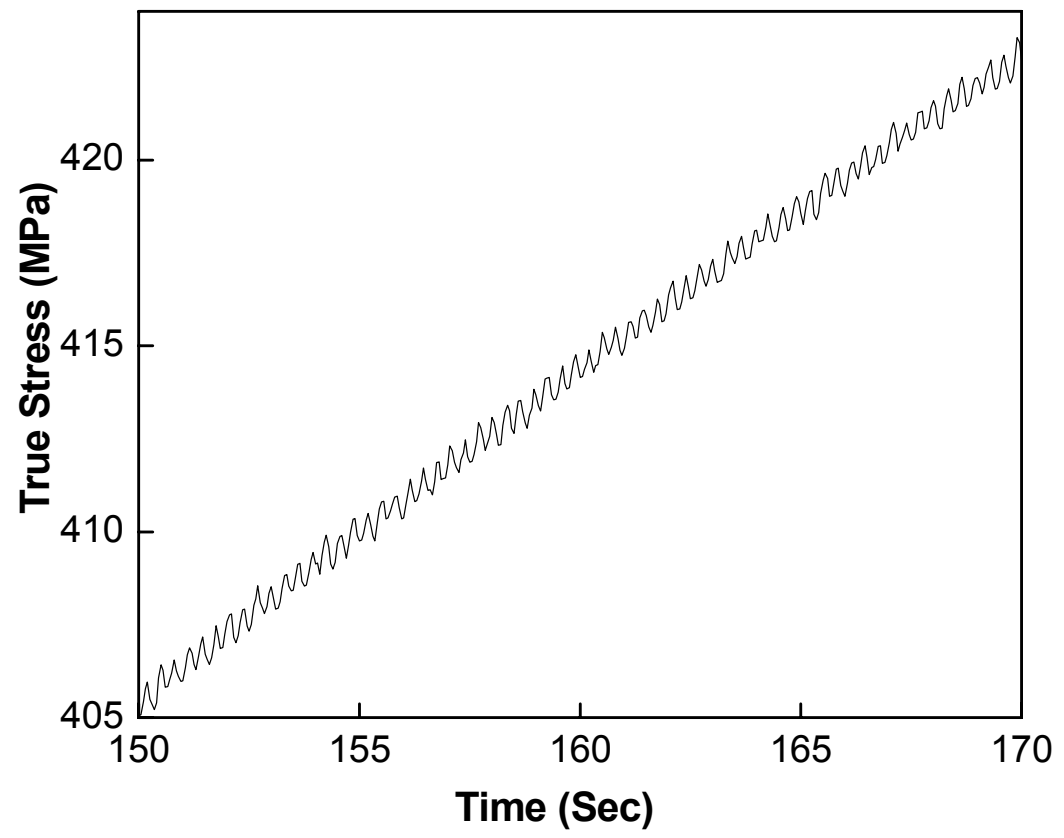

**Figure 1:** Typical segments of the True Stress vs Time curve for (a) Al 2.5%Mg alloy and (b)Low Carbon Steel sample deformed at strain rate of $3.85\times10^{-4}$ $Sec^{-1}$ and $1.12\times10^{-3}$ $Sec^{-1}$ respectively.

## 3. Methods of Analysis

Scale invariance has been found to hold empirically for a number of complex systems, and the correct evaluation of the scaling exponents is of fundamental importance in assessing if universality classes exist [16]. We make use of two complementary scaling analysis methods the standard deviation analysis (SDA) and the diffusion entropy analysis (DEA) to study the scaling behavior of the PLC effect [17-20].

*SDA and DEA*

SDA and DEA together are found very efficient to detect the exact scaling behavior of a time series. The need for using these two methods to analyze the scaling properties of a time series is to discriminate the stochastic nature of the data: Gaussian or Levy [18]. Recently, Scafetta *et al.* [21] had shown that to distinguish between fractal Gaussian intermittent noise and Levy-walk intermittent noise, the scaling results obtained using DEA should be compared with that obtained from SDA.

These methods are based on the prescription that numbers in a time series $\{\xi_i\}$ are the fluctuations of a diffusion trajectory; see Refs. [17-20] for details. Therefore, we shift our attention from the time series $\{\xi_i\}$ to the probability density function (pdf) $p(x,t)$ of the corresponding diffusion process. Here $x$ denotes the variable collecting the fluctuations and is referred to as the diffusion variable. The scaling property of $p(x,t)$ takes the form

$$p(x,t) = \frac{1}{t^{\delta}} F\left(\frac{x}{t^{\delta}}\right) \qquad (1)$$

In the SDA one examines the scaling properties of the second moment of the diffusion process generated by a time series. SDA is based on the evaluation of the standard deviation $D(t)$ of the variable x, and yields

$$D(t) = \sqrt{\langle x^2;t\rangle - \langle x;t\rangle^2} \propto t^{\gamma} \qquad (2)$$

The exponent $\gamma$ is interpreted as the scaling exponent. It is evaluated from the gradient of the fitted straight line in the log-log plot of $D(t)$ against $t$.

The DEA was developed [17] as an efficient way to detect the scaling and memory in time series for variables in complex systems. This procedure has been successfully applied to sociological, astrophysical and biological time series. DEA focuses on the scaling exponent δ evaluated through the Shannon entropy *S(t)* of the diffusion generated by the fluctuations $\{\xi_i\}$ of the time series [17,18]. Here, the pdf of the diffusion process, $p(x,t)$, is evaluated by means of the sub trajectories $x_n(t) = \sum_{i=0}^{t} \xi_{i+n}$ with $n$ =0,1,.. If the scaling condition of Eq. (1) holds true, it is easy to prove that the entropy

$$S(t) = -\int_{-\infty}^{\infty} p(x,t)\ln[p(x,t)]dx \qquad (3)$$

increases in time as

$$S(t) = A + \delta\ln(t) \qquad (4)$$

with

$$A = -\int_{-\infty}^{\infty} dy F(y)\ln[F(y)] = \text{Constant}, \qquad (5)$$

where $y = \frac{x}{t^{\delta}}$. Eq. (4) indicates that in the case of a diffusion process with a scaling pdf, its entropy $S(t)$ increases linearly with $\ln(t)$. The scaling exponent $\delta$ is evaluated from the gradient of the fitted straight line in the linear-log plot of $S(t)$ against $t$.

Finally we compare $\gamma$ and $\delta$. For fractional Brownian motion the scaling exponent $\delta$ coincides with the $\gamma$ [19]. For random noise with finite variance, the diffusion distribution $p(x,t)$ will converge, according to the central limit theorem, to a Gaussian distribution with $\gamma = \delta = 0.5$. If $\gamma \neq \delta$ the scaling represents anomalous behavior. The diffusion processes characterized by Levy flights fall into the class of anomalous diffusion. In this case the Eq. (1) still holds true but the variance is not finite and, therefore, the variance scaling exponent cannot be defined. Another interesting example of the anomalous diffusion is the case of Levy-walk, which is obtained by generalizing the central limit theorem. In this particular kind of diffusion process the second moment is finite but the scaling exponents $\gamma$ and $\delta$ are found to obey the relation

$$\delta = \frac{1}{3 - 2\gamma} \qquad (6)$$

[22], instead of being equal.

## 4. Results and Discussions

Fig. 2 and Fig. 3 show the typical plot of the SDA and the DEA for Al-2.5%Mg alloy and low carbon steel respectively. The scaling exponents are shown in the figures. It is seen that the SDA scaling exponent ($\gamma$) and the DEA scaling exponent ($\delta$) are high for all the cases. The high values of the exponents imply a strong persistence of the stress fluctuations. Moreover, we note that the SDA scaling exponents $\gamma$ are larger than the DEA scaling exponents $\delta$ and obey the Levy walk scaling relation (6) for both the alloys in the entire span of strain rate. These results suggest that the Levy walk scaling properties emerge from the local strain rate anomalies which in turn are related to the intermittency of the dislocation activity.

From the scaling analysis it is clear that the PLC effect follows a Levy walk scaling behavior irrespective of the alloy type, microstructure and the crystal structure. Now the question arises is- how robust is the observed scaling property? To check the robustness of the observed scaling behavior we have corrupted 2% of the stress data (randomly chosen) by adding noise of magnitude multiple of the standard deviation (std) of the stress data. We found that addition of noise of magnitude of four times of the std the scaling exponents did not change and the scaling behavior is retained by an addition of noise of magnitude of ten times of std. This confirms the robustness of the scaling property of the PLC effect.

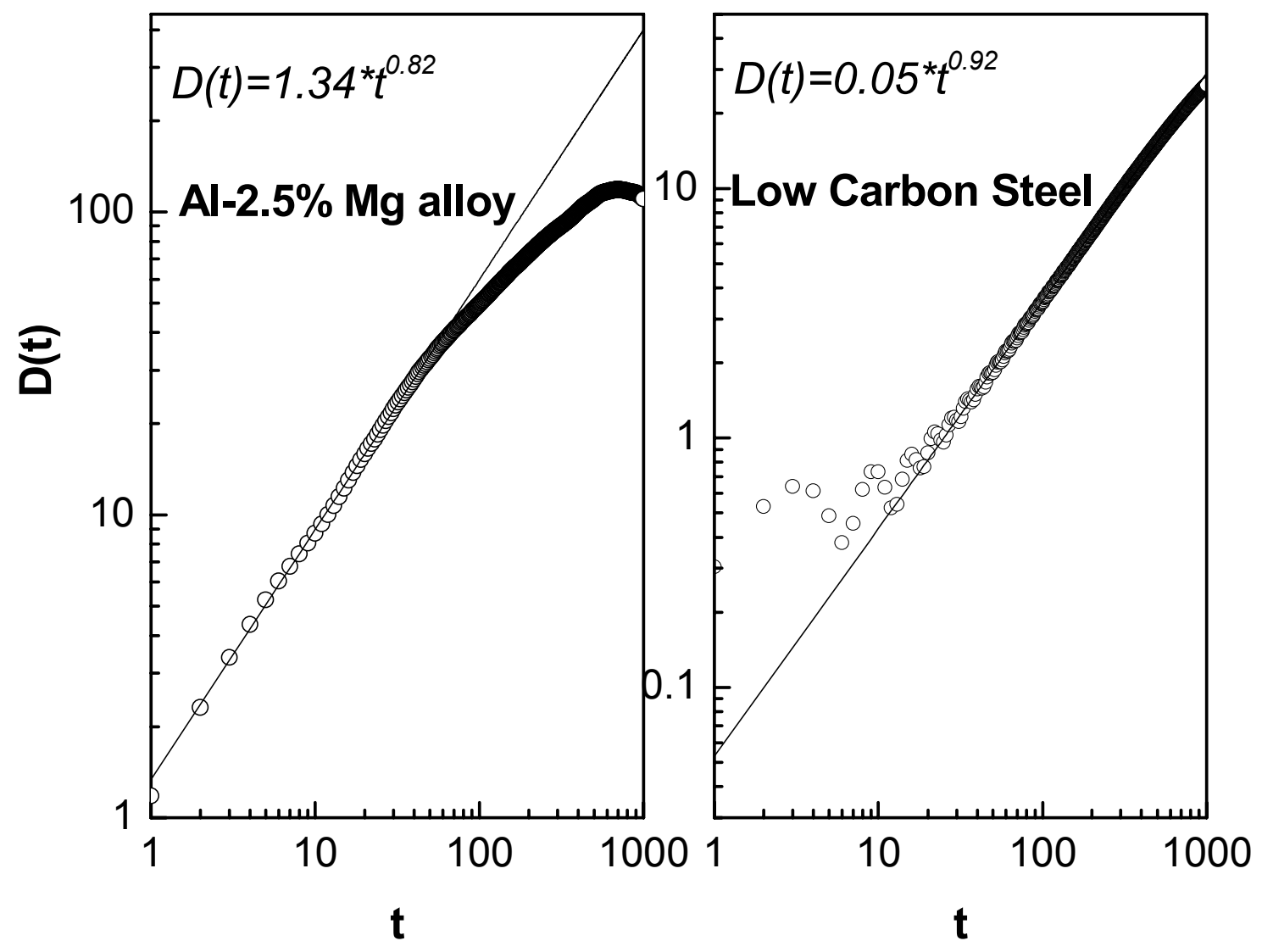

**Figure 2:** SDA of the Stress vs. Time data obtained from Al-2.5%Mg alloy and Low carbon steel samples during tensile deformation at a strain rate of $3.85\times10^{-4}$ $Sec^{-1}$ and $1.12\times10^{-3}$ $Sec^{-1}$ respectively.

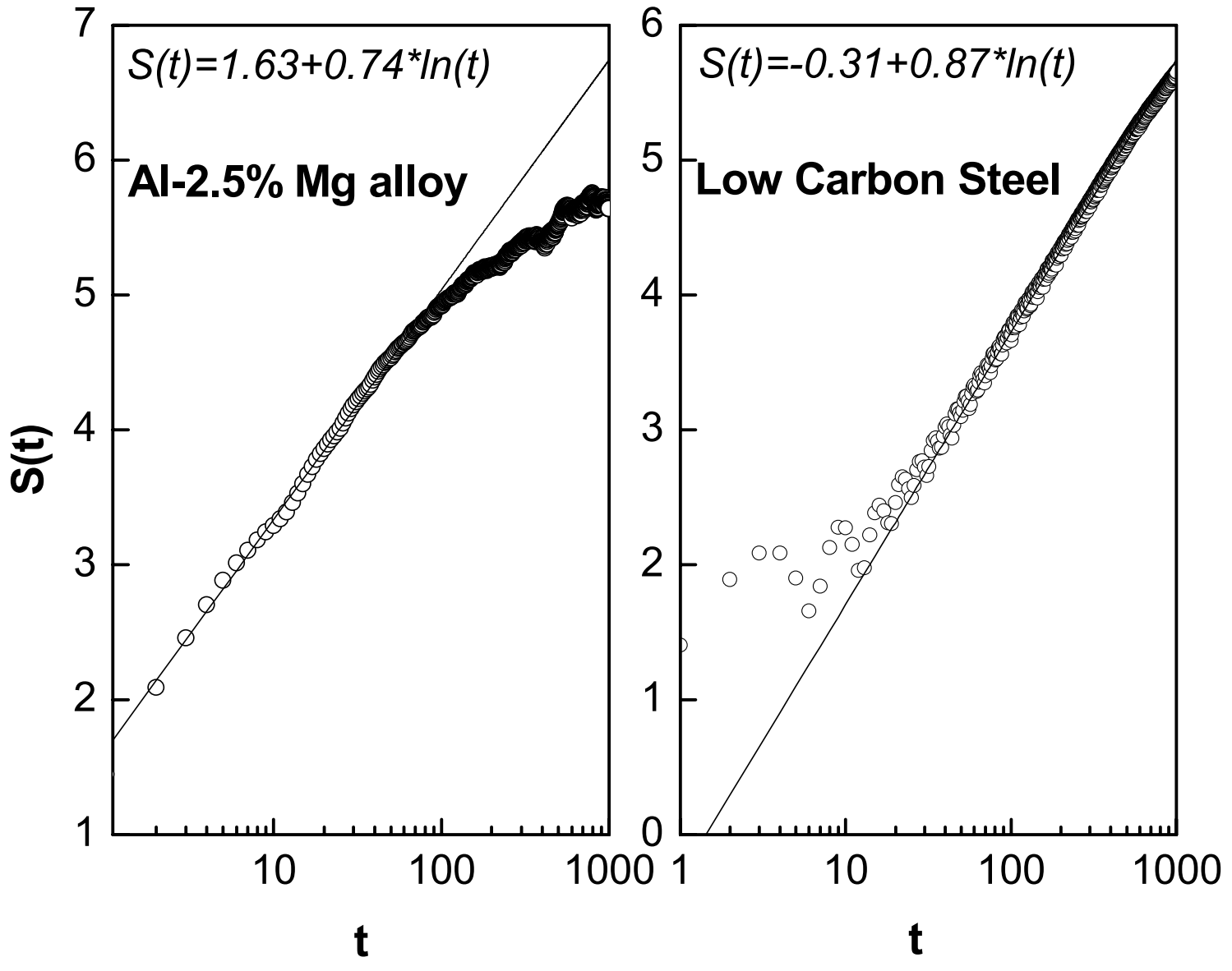

**Figure 3:** DEA of the Stress vs. Time data obtained from Al-2.5%Mg alloy and Low carbon steel samples during tensile deformation at a strain rate of $3.85\times10^{-4}$ $Sec^{-1}$ and $1.12\times10^{-3}$ $Sec^{-1}$ respectively.

Alloys studied in this work are entirely different as far as the crystal structure and the microstructure are concerned. The Al-2.5%Mg possesses face centered cubic (f.c.c) crystal structure and it is a substitutional alloy. On the contrary the low carbon steel possesses body centered cubic (b.c.c) crystal structure with interstitial solute atoms. But the scaling analysis on the PLC effect of both the alloys reveals the persistence Levy walk type of scaling behavior. This establishes the universality of the scaling behavior of PLC effect. The universality of the scaling behavior suggests that though types of ingredient changes the underlying reason behind the PLC effect i.e. the jerky movement of dislocations remains almost similar.

**5. Conclusions**

We have studied the scaling behavior of the Portevin-Le Chatelier effect in different alloy systems using two complementary scaling analysis methods: SDA and DEA. The analyses were performed for a wide range of strain rates where different types of deformation bands are observed. The relation of the two scaling exponents in each strain rates obtained through our analysis for both the alloys clearly suggests that PLC effect follow a Levy walk type of scaling behavior, independent of the material and strain rate. The scaling behavior is also found to be a robust one.

## References

[1] J. Balik, P. Lukac, *Portevin-Le Chatelier instabilities in Al-3Mg conditioned by strain rate and strain*, Acta Metall. Mater. **41** (1993) 1447.

[2] P. G. McCormick, *The ageing time for serrated yielding in a low carbon steel*, Scripta Metal. **7** (1973) 945.

[3] A. H. Cottrel, *A note on the Portevin- Le Chatelier effect*, Phil. Mag. **44** (1953) 829.

[4] A. W. Sleeswyk, *Slow strain hardening of ingot iron*, Acta Metal. **6** (1958) 598.

[5] A. Van den Beukel, *Theory of the effect of dynamic strain ageing on mechanical properties*, Physica Status Solidi(a) **30** (1975) 197.

[6] M. Zaiser, P. Hahner, *Oscillatory modes of plastic deformation: theoretical concepts*, Physica Status Solidi(b) **199** (1997) 267.

[7] L. P. Kubin, C. Fressengeas, G. Ananthakrishna, *Dislocation in Solids*, edited by F.R.N. Nabaro, M.S. Duesbery, Elsevier Science, Amsterdam **11** (2002) 101.

[8] K. Chihab, Y. Estrin, L. P. Kubin, J. Vergnol, *The kinetics of the Portevin- Le Chatelier bands in an Al-5at% Mg alloy*, Scripta Metal. **21** (1987) 203.

[9] P. Hahner, E. Rizzi, *On the kinematics of Portevin- Le Chatelier bands: theoretical and numerical modelling*, Acta Mater. **51** (2003) 3385.

[10] M. S. Bharathi, M. Lebyodkin, G. Ananthakrishna, C. Fressengeas and L. P. Kubin, *The hidden order behind jerky flow*, Acta Mater. **50** (2002) 2813.

[11] B. B. Mandelbrot, *The Fractal Geometry of Nature*, Freeman, New York 1983.

[12] P. Bak, K. Christensen, L. Danon, T. Scanlon, *Unified Scalinf Laws for Earthquakes*, Phys. Rev. Lett. **88** (2002) 178501.

[13] A. P. Seibesma, H.J.J. Jonker, *Anomalous Scaling of Cumulus Cloud Boudaries*, Phys. Rev. Lett. **85** (2000) 214.

[14] S.N. Dorogovtsev, J.F.F. Mendes, *Scaling Properties of Scale-free evolving networks: Continuous approach*, Phys. Rev. E **63** (2001) 056125.

[15] A. Sarkar, P. Barat, *Scaling Analysis on Indian Foreign Exchange Market*, In Press Physica A.

[16] H. E. Stanley, L. A. N. Amaral, P. Gopikrishnan, P. Ch. Ivanov, T.H. Keitt, V. Plerou, *Scale invariance and universality: organizing principles in complex systems*, Physica A **281** (2000) 60.

[17] N. Scafetta, P. Hamilton, P. Grigolini, *The thermodynamics of social process: the teen birth phenomenon*, Fractals **9** (2001) 193.
[18] N. Scafetta, P. Grigolini, *Scaling detection in time series: diffusion entropy analysis*, Phys. Rev. E **66** (2002) 036130.
[19] N. Scafetta, V. Latora, P. Grigolini, *Lévy statistics in coding and non-coding nucleotide sequences*, Phys. Lett. A **299** (2002) 565.
[20] N. Scafetta, B. J. West, *Solar flare intermittency and Earth temperature anomalies*, Phys. Rev. Lett. **90** (2003) 248701.
[21] N. Scafetta, B. J. West, *Multi-scaling comparative analysis of time series and a discussion on 'earthquake conversations' in California*, Phys. Rev. Lett. **92** (2004) 138501.
[22] N. Scafetta, *An entropic approach to the analysis of time series,* Phd Thesis, University of North Texas.